\documentclass[a4paper]{jpconf}
\usepackage{graphicx}

\newcommand{\siiv}{Si\,{\sc iv}}

\newcommand{\civ}{C\,{\sc iv}}

\newcommand{\aliii}{Al\,{\sc iii}}

\newcommand{\mgii}{Mg\,{\sc ii}}

\begin{document}
\title{The central structure of Broad Absorption Line QSOs: observational characteristics in the cm-mm wavelength domain}

\author{G Bruni$^{1,2,3}$, K-H Mack$^1$, 
D Dallacasa$^{1,2}$, 
F M Montenegro-Montes$^4$,
C R Benn$^5$,
R Carballo$^6$,
J I Gonz\'alez-Serrano$^7$,
J Holt$^8$ and
F Jim\'enez-Luj\'an$^{3,5,7}$}
\vspace{0.3cm}
\address{$^1$INAF-Istituto di Radioastronomia, via Piero Gobetti, 101, I-40129 Bologna, Italy}
\address{$^2$Universit\`a di Bologna, Dip. di Astronomia, via Ranzani, 1, I-40127 Bologna, Italy}
\address{$^3$Dpto. de F\'isica Moderna, Univ. de Cantabria, Avda de los Castros s/n, E-39005 Santander, Spain}
\address{$^4$European Southern Observatory, Alonso de C\'ordova 3107, Vitacura, Casilla 19001, Santiago, Chile}
\address{$^5$Isaac Newton Group, Apartado 321, E-38700 Santa Cruz de La Palma, Spain}
\address{$^6$Dpto. de Matem\'atica Aplicada y Ciencias de la Computaci\'on, Univ. de Cantabria, ETS Ingenieros de Caminos, Canales y Puertos, Avda de los Castros s/n, E-39005 Santander, Spain}
\address{$^7$Instituto de F\'isica de Cantabria (CSIC-Universidad de Cantabria), Avda. de los Castros s/n, E-39005 Santander, Spain}
\address{$^8$Leiden Observatory, Leiden University, P.O. Box 9513, NL-2300 RA Leiden, The Netherlands}

\ead{bruni@ira.inaf.it}


\begin{abstract}
Accounting for $\sim$20\% of the total QSO population, Broad Absorption Line QSOs are still an unsolved problem in the AGN context. They present wide troughs in the UV spectrum, due to material with velocities up to 0.2 c toward the observer. The two models proposed in literature try to explain them as a particular phase of the evolution of QSOs or as normal QSOs, but seen from a particular line of sight.\\
We built a statistically complete sample of Radio-Loud BAL QSOs, and carried out an observing campaign to piece together the whole spectrum in the cm wavelength domain, and highlight all the possible differences with respect to a comparison sample of Radio-Loud non-BAL QSOs. VLBI observations at high angular resolution have been performed, to study the pc-scale morphology of these objects. Finally, we tried to detect a possible dust component with observations at mm-wavelengths.\\
Results do not seem to indicate a young age for all BAL QSOs. Instead a variety of orientations and morphologies have been found, constraining the outflows foreseen by the orientation model to have different possible angles with respect to the jet axis.    
\end{abstract}


\section{Introduction}
\vspace{0.3cm}
Broad Absorption Line (BAL) QSOs are a particular and not yet understood class of AGN: their spectra show wide troughs towards the blue wing of some UV emission lines (\mgii, \aliii, \siiv, \civ), due to ionized gas
with outflow velocities up to 0.2 c. They account for $\sim$20\% of the QSO population. Two main scenarios have been considered to explain their nature: the \emph{orientation} model and the \emph{evolutionary} one. The former, proposed by \cite{Elvis}, foresees the BAL outflows to be present in all QSOs, but visible only when intercepting the line of sight of the observer, thus a particular orientation can be supposed for these objects. The latter model (\cite{Briggs, Lipari}) explains the BAL phenomenon as a particular phase in the evolutionary sequence of QSOs, during which a dusty shell is being expelled from the AGN. In this case a greater amount of dust emission should be present with respect to \emph{normal} QSOs. A third model from Punsly (\cite{Punsly1,Punsly2}) proposes that the BAL phenomenon is due to polar winds above the inner jet, thus would be visible only for lines of sight close to the jet axis.  

In this framework the radio emission is an additional diagnostic tool to test the proposed models. We selected a sample of Radio-Loud BAL QSOs, cross-correlating the fourth edition of the SDSS Quasar Catalogue (\cite{Schneider}) with the FIRST catalogue (\cite{Becker}), with a constraint of $S_{1.4~\rm{GHz}}>30$ mJy for the flux density at 1.4 GHz and $1.7<z<4.7$ for the redshift, in order to have sufficient flux density for VLBI studies and to allow the classification as BAL QSO through the CIV and MgII lines in the optical band. The whole sample of 25 objects, together with a comparison sample of 34 non-BAL QSOs, is presented in \cite{Bruni}. 


\section{Radio Observations}
\vspace{0.3cm}
In 2009-2010 we carried out an observational campaign to cover the cm-wavelength spectrum of the two BAL and non-BAL QSO samples, using both single dish (100-m Effelsberg) and interferometers (GMRT, VLA) in the frequency range from 240 MHz to 43 GHz. Our target was the investigation of the differences between the two samples, aiming at determining which of the three models is the most probable scenario for BAL QSOs. There are three observables that can give clues about that: (1) the spectral index of the synchrotron spectrum is a statistical indicator of the orientation of the source (\cite{Orr}), since flatter spectral indices imply lines of sight closer to the jet axis. If a particular orientation was present in BAL QSOs, the spectral index distribution for the sample would be different with respect to the non-BAL sample; (2) the synchrotron peak frequency is related to the age of the source, younger sources having a smaller linear size and a higher peak frequency (\cite{ODea}). If BAL QSOs were young objects in a particular evolutionary phase, the ratio of Gigahertz-Peaked Spectrum (GPS) sources would be greater than in the non-BAL sample; (3) polarization and Rotation Measure (RM) are indicative of the magnetic field strength and particle density around the central environment of the AGN. Finally, VLBI studies can help in determining the morphology and orientation of these particular class of QSOs. 


\subsection{Spectra and polarization}
We fitted the radio spectra of the sources through $\chi^2$ minimization, with a power law and a parabola in the $log{S_\nu} - log{\nu}$ plane, in order to determine whether a synchrotron peak frequency in the GigaHertz range was present. In some cases a low-frequency component was found, likely corresponding to old components, and not indicating a young age for the source. We found 9 BAL QSOs and 8 non-BAL QSOs showing a GPS, and thus a similar fraction of GPS sources in the two samples (36\% and 24\%), but further observations are needed to confirm their classification, since the angular resolution of the VLA in C configuration only allowed us to give an upper limit of 20 kpc to the linear size in the 8 GHz maps, while the maximum size for a GPS source is 1 kpc. Fits and spectra of the two samples can be found in \cite{Bruni}.

\begin{figure}
\begin{center}
\includegraphics[width=165mm]{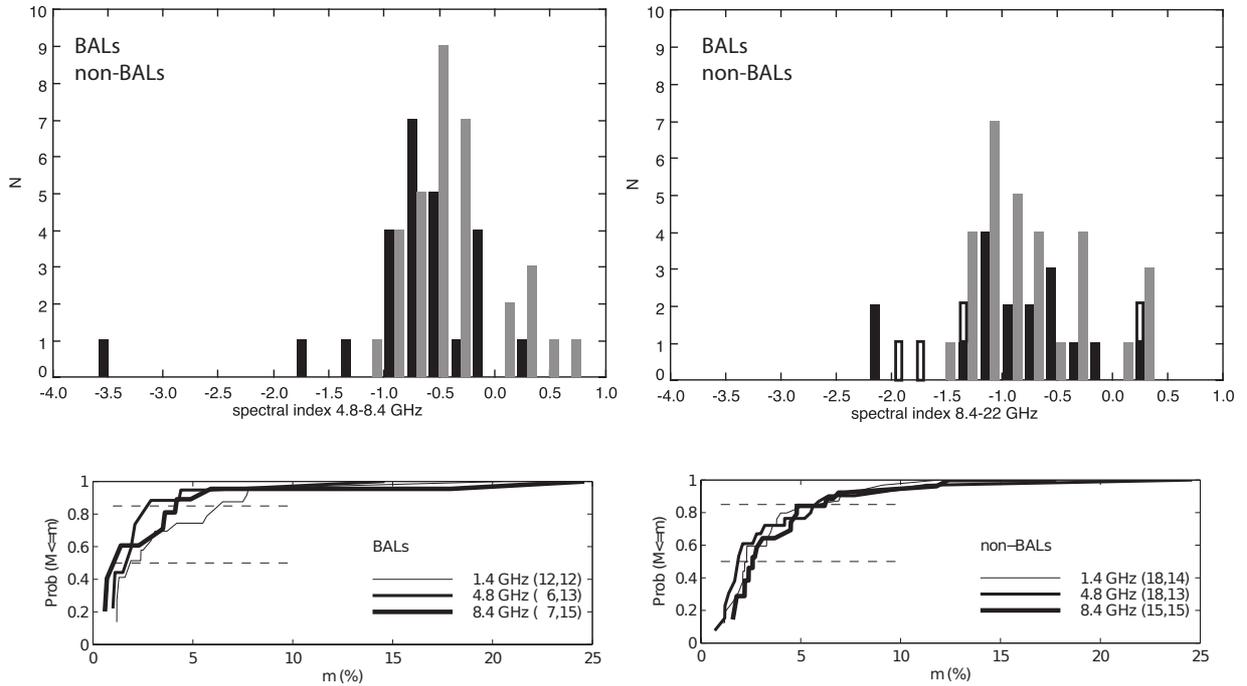}
\end{center}
\caption{\label{polarization} Top panels: distribution of the radio spectral indices ($\alpha_{4.8}^{8.4}$ and $\alpha_{8.4}^{22}$) for the 
BAL (black) and non-BAL (grey) QSO samples. Upper limits for BAL QSOs are in white.
Bottom panels: cumulative distribution of the fractional polarization for the two samples at three frequencies. The pair of numbers in parenthesis indicates the number of detections and the number of upper limits for each frequency. The dashed lines show the 50\% percentile (i.e. the median) and the 85\% percentile.}
\end{figure}

The spectral index (defined as $\alpha_{1,2}=[(\log{S_2}-\log{S_1})/(\log{\nu_2}-\log{\nu_1})]$, where $\nu_2>\nu_1$) was calculated for all sources in the frequency interval 4.8-8.4 GHz and 8.4-22 GHz, i.e. well above the peak frequency for all of them. A variety of spectral indices have been found, both flat ($\alpha<0.5$) and steep ($\alpha>0.5$), suggesting a wide range of orientations for BAL QSOs, similar to the non-BAL QSO sample (see Fig.~\ref{polarization}). A Kolmogorov-Smirnov test comparing the two spectral index distributions $\alpha_{4.8}^{8.4}$ and $\alpha_{8.4}^{22}$ for BAL and non-BAL QSOs gives a low significance level (74\% and 0.3\%) for the two to be different. Moreover, the same test gives a significative evidence that BAL QSOs are not flatter than non-BAL QSOs, in contradiction to the hypothesis from \cite{Punsly1,Punsly2}.

Measurements or upper limits of the fractional polarization ($m$) have been obtained for the two samples, and a  Kaplan-Meier estimator was used to compare the two, showing that the cumulative distributions are very similar (see Fig. \ref{polarization}). The median value for $m$ is in the range 1.8-2.5\% for both samples. We calculated the RMs for all the sources with at least 3 values of $m$ at different frequencies (4 BAL QSOs and 10 non-BAL QSOs): they show similar values despite the small statistics, the only outlier being BAL QSO 1624+37, with the second highest value of RM among extra-galactic sources (RM=$-$18350$\pm$570 rad m$^{-2}$, \cite{Benn}).

\subsection{Morphology and orientation}
In 2009-2010 we obtained also high-resolution images for 11 BAL QSOs in VLBI technique with the EVN (5 GHz) and VLBA (5, 8.4 GHz) arrays (\cite{Bruni2}). 9 of 11 sources present a resolved structure (82\%), and various morphologies are visible (see Fig. \ref{VLBI}). Double (0756+37, 1102+11), core-jet (0044+00, 0816+48, 0849+27, 1304+13) and symmetric (1014+05, 1237+47, 1603+30) structures have been found, so different orientations can be argued. Unresolved sources could be beamed jets towards the observer or extremely young sources, like 1406+34. Projected linear sizes are constrained under 1 kpc: this confirms the GPS classification for 0756+37, 1237+47 and 1603+30.

\begin{figure}
\begin{center}
\includegraphics[width=160mm]{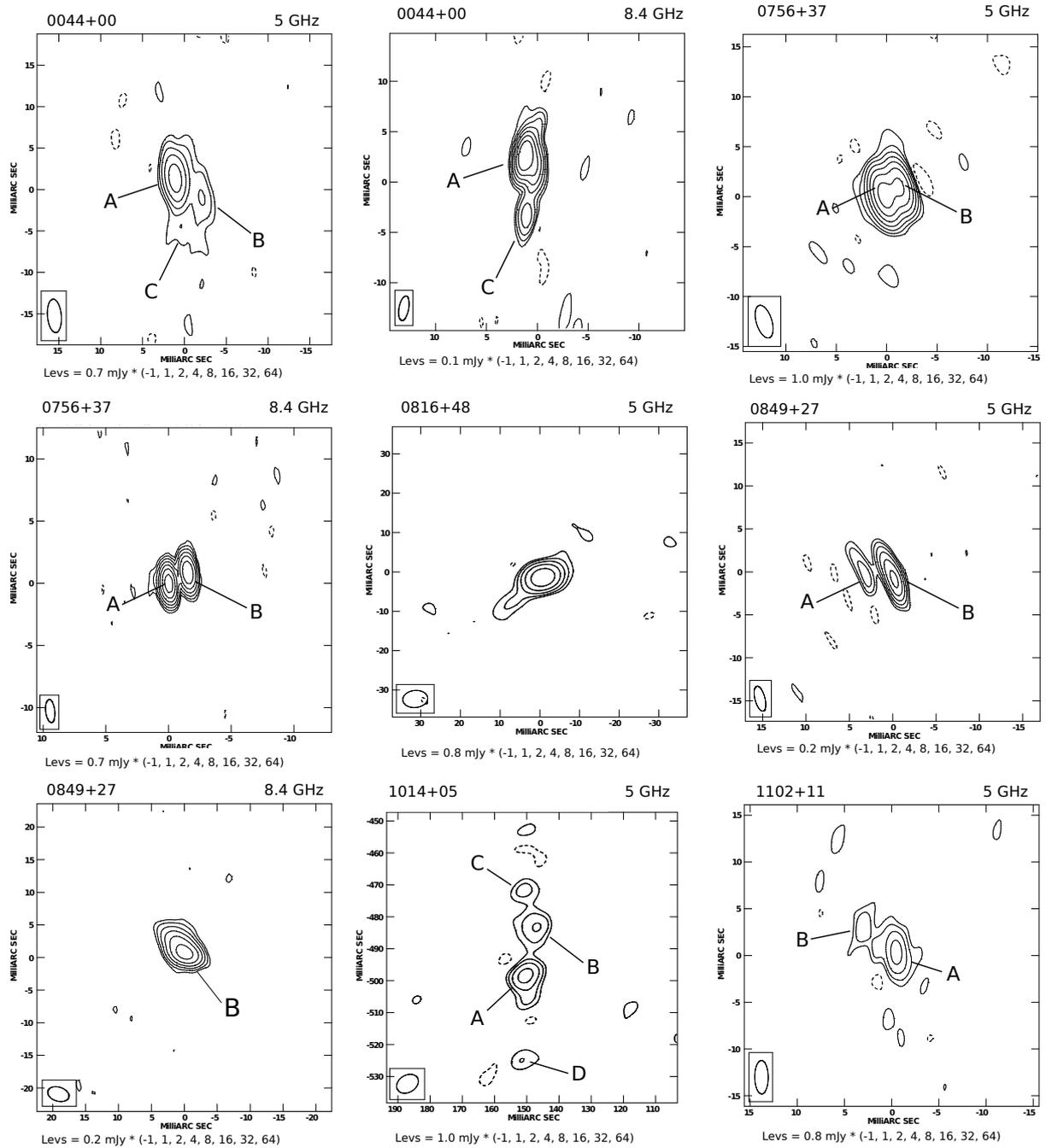}
\end{center}
\caption{\label{VLBI} High angular resolution maps of 11 BAL QSOs, obtained with the VLBA and the EVN.}
\end{figure}

\addtocounter{figure}{-1}
\begin{figure}
\begin{center}
\includegraphics[width=160mm]{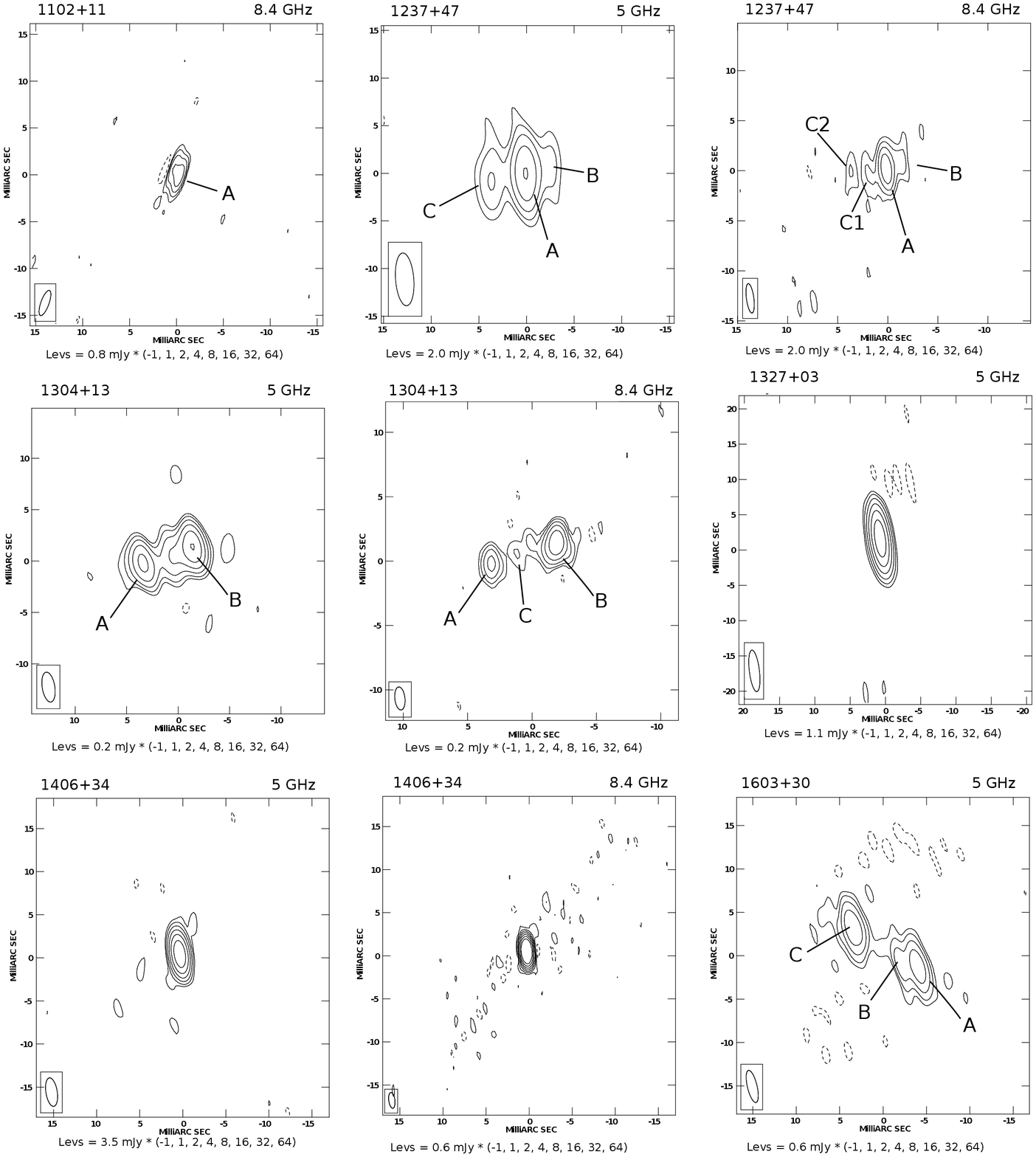}
\end{center}
\caption{\label{VLBI} Continued.}
\end{figure}

Four BAL QSOs were already resolved in VLA maps, showing 1 elongated (1103+11) and 3 core-lobes (0816+48, 0849+27, 1603+30) morphologies, with projected linear sizes between 20 and 200 kpc.


\section{Dust detection}
\begin{figure}
\begin{center}
\includegraphics[width=165mm]{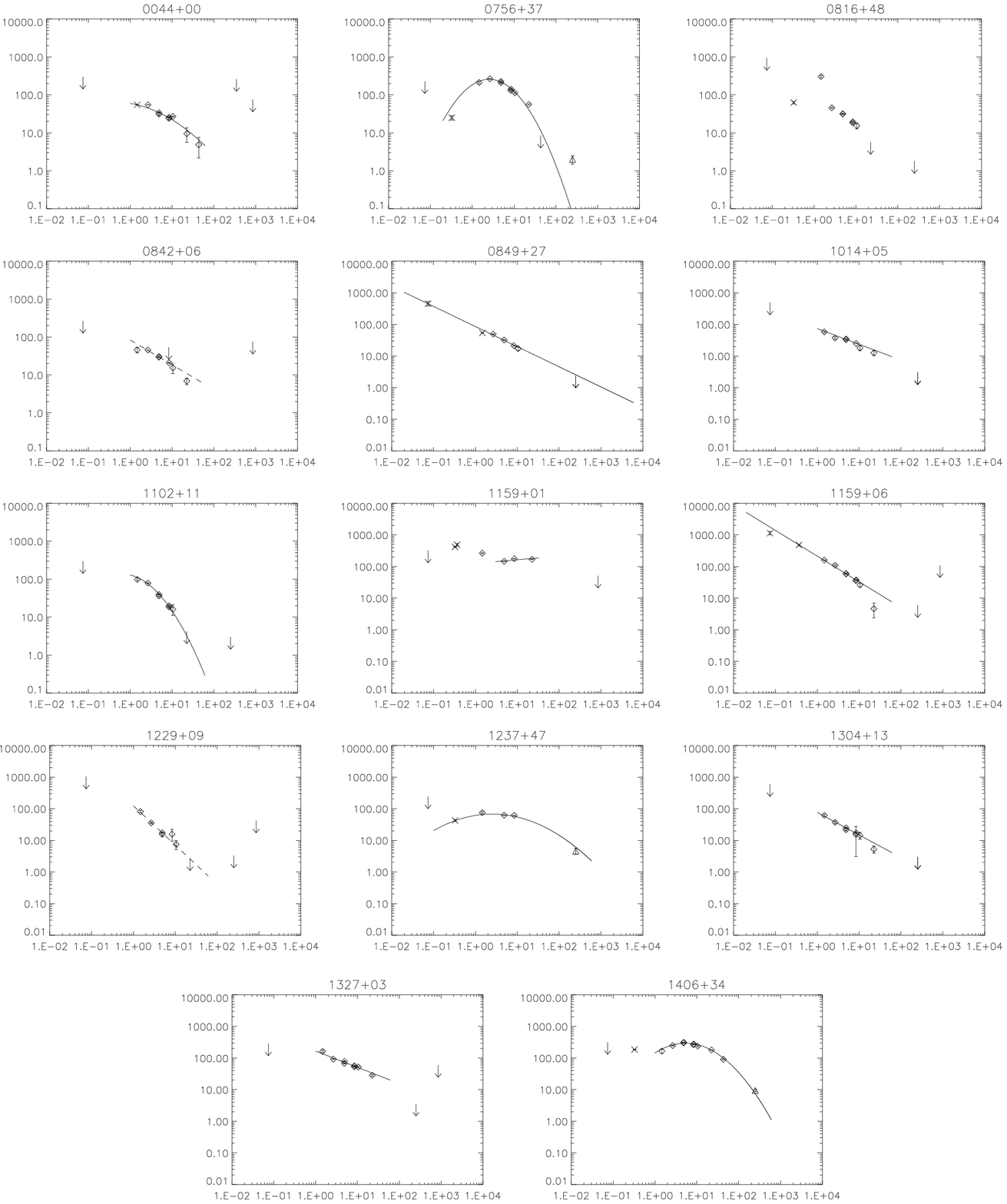}
\end{center}
\caption{Spectra of 14 BAL QSOs observed in the mm-band (x-axis: GHz; y-axis: mJy). 250 GHz flux densities from IRAM-30m, 345 and 850 GHz from APEX (triangles). Flux densities at lower frequencies are taken from \cite{Bruni} (crosses, $\nu<1.4$ GHz: literature; rhombi, $\nu\ge1.4$ GHz: our observational campaign).}
\label{dust} 
\end{figure}

\vspace{0.3cm}
In the mm-wavelength domain we performed observations with the IRAM-30m single dish (250 GHz) and with APEX (345 and 850 GHz), to detect the grey-body dust emission in the BAL QSO sample. The amount of dust is one of the main differences in the two scenarios proposed to explain the nature of BAL QSOs, since a young source, still expelling its dust cocoon, should present an excess in the mm-band emission (in the redshift range of our sample). The two observations were planned to detect the peak (850 GHz) and the power-law slope (240, 345 GHz) of this emission. We found 1 out of 14 observed sources ($\sim7\%$) showing a clear dust component not correlated with the synchrotron emission (0756+37, see Fig. \ref{dust}).  

For this study we refer, as a comparison, to the dust emission study published by Omont et al. for QSOs with z$\sim$2 (\cite{Omont}): they found that $\sim25\%$ of their sample shows hints of dust at 250 GHz.
Since their study has been carried out with the same instrument, and obtaining a similar RMS, we can safely compare with this result, concluding that Radio-Loud BAL QSOs does not seem to present a larger fraction of dust-rich objects with respect to the QSO population.

More observations, already scheduled at APEX, will increase the statistic and thus the reliability of these preliminary results.

\section{Conclusions}
\vspace{0.5cm}

We can summarize our findings as follows:
\begin{itemize}
 \item A variety of orientation for BAL QSOs have been found,
both from spectral index and from VLBI morphology studies.
\item Radio spectra do not seem to indicate a young age, since the 
fraction of GPS sources is similar to non-BAL QSOs. Moreover, in some cases 
a low-frequency emission is present, suggesting an old radio component.
\item VLBI maps for 11 BAL QSOs show different morphologies and projected linear sizes under 1 kpc. Bigger linear sizes are
not excluded for BAL QSOs, since 4 sources were already resolved with the VLA, resulting in projected linear sizes between 20 and 200 kpc.  
\item Radio-Loud BAL QSOs do not seem to be dust-rich, since preliminary results only show $\sim7\%$ of the sources with a flux density excess in the mm-band, comparable with previous findings for the QSO population.
\end{itemize}

Excluding a young age for Radio-Loud BAL QSOs, the most probable scenario seems to be the one proposed by \cite{Elvis},
but only if it can account for outflows with a variety of angles with respect to the jet axis, able to justify the different orientations found in radio observations.

\subsection*{Acknowledgments}
\small{
We are grateful to the AHAR11 LOC for the economical support.\\
Part of this work was supported by a grant of the Italian Programme for
Research of Relevant National Interest (PRIN No. 18/2007, PI: K.-H. Mack)
The authors acknowledge financial support from the Spanish Ministerio de Ciencia 
e Innovaci\'on under project AYA2008-06311-C02-02.\\
This work has benefited from research funding from the European Union's 
sixth Framework Programme under RadioNet grant agreement no. 227290.\\ 
This work has been partially based on observations with the 100-m 
telescope of the MPIfR (Max-Planck-Institut f\"ur Radioastronomie) at Effelsberg.\\ 
The National Radio Astronomy Observatory is a facility of the National Science Foundation operated under cooperative 
agreement by Associated Universities, Inc.\\ 
The European VLBI Network is a joint facility of European, Chinese, South African and other radio astronomy institutes funded by their national research councils.\\
This work has been partially based on observations carried out with the IRAM 30m single-dish. IRAM is supported by INSU/CNRS (France), MPG (Germany) and IGN (Spain).\\
This publication is partially based on data acquired with the Atacama Pathfinder Experiment (APEX). APEX is a collaboration between the Max-Planck-Institut fur Radioastronomie, the European Southern Observatory, and the Onsala Space Observatory.\\
This research has made use of the NASA/IPAC Infrared Science Archive and NASA/IPAC Extragalactic Database (NED) 
which are both operated by the Jet Propulsion Laboratory, 
California Institute of Technology, under contract with the National Aeronautics and Space Administration.\\ 
Use has been made of the Sloan Digital Sky Survey (SDSS) Archive. The SDSS is managed by the Astrophysical Research Consortium (ARC) for the participating institutions: The University of Chicago, Fermilab, the Institute for Advanced Study, the Japan Participation
Group, The John Hopkins University, Los Alamos National Laboratory, the Max-Planck-Institute for Astronomy (MPIA), the 
Max-Planck-Institute for Astrophysics (MPA), New Mexico State University, University of Pittsburgh, Princeton University,
the United States Naval Observatory, and the University of Washington.\\ 
} 



\end{document}